\documentclass[rnote]{aa} % for the research notes
\usepackage{graphicx}

\usepackage{epstopdf}
\usepackage{natbib}

\usepackage{txfonts}
\def\4he{$^4$He}

\def\msol{M_\odot}

\def\sec#1{Sec. \ref{#1}}

\def\h2{$\mathrm{H}_2$}
\def\nh3{$\mathrm{NH}_3$}
\def\mic{\mathrm{ }\mu\mathrm{m}}

\def\citeapos#1{\citeauthor{#1}'s (\citeyear{#1})}  %Defines a possessive citation

\begin{document}
   \title{Classifying the secondary component of the binary star W Aquilae}

   %\subtitle{Paper I: Initial results from the NOT}

   \author{T. Danilovich
          \inst{1}
          \and
          G. Olofsson \inst{2}
          \and
          J. H. Black \inst{1}
          \and
          K. Justtanont \inst{1}
          \and
          H. Olofsson \inst{1}
          }

   \institute{Department of Earth and Space Sciences, Chalmers University of Technology, Onsala Space Observatory, SE-439 92 Onsala, Sweden
              \and 
              Department of Astronomy, Stockholm University, AlbaNova University Center, SE-106 91 Stockholm, Sweden
             \\ \email{taissa@chalmers.se}}

   \date{Received 19 February 2014; accepted 23 December 2014}

% \abstract{}{}{}{}{} 
% 5 {} token are mandatory
 
  \abstract
   {}
   {The object W\,Aql is an asymptotic giant branch (AGB) star with a faint companion. By determining more carefully the properties of the companion, we hope to better constrain the properties of the AGB star.}
   {We present new spectral observations of the binary star W\,Aql at minimum and maximum brightness and new photometric observations of W\,Aql at minimum brightness.}
   {The composite spectrum near minimum light is predominantly from the companion at wavelengths $\lambda < 6000 \,\AA$. This spectrum can be classified as F8 to G0, and the brightness of the companion is that of a dwarf star. Therefore, it can be concluded that the companion is a main sequence star. From this, we are able to constrain the mass of the AGB component to 1.04 -- 3 $\msol$ and the mass of the W\,Aql system to 2.1 -- 4.1 $\msol$. Our photometric results are broadly consistent with this classification and suggest that the main sequence component suffers from approximately 2 mag of extinction in the V band primarily due to the dust surrounding the AGB component.}
   {}


   \maketitle
%
%________________________________________________________________

\section{Introduction}

It has long been known that the S-type asymptotic giant branch (AGB) star W\,Aql has a faint companion. \cite{Herbig1965} noted that at minimum brightness W\,Aql took on spectral features similar to an F5 or F8 star, thus first identifying it as a spectroscopic binary. High resolution Hubble Space Telescope (HST) images have since shown that the binary pair is separated by 0\farcs46 \citep{Ramstedt2011}. The object W Aql is a Mira variable with a period of 490 days and an amplitude of approximately 7 mag, ranging from 7.3 to 14.3 mag in the V band. Due to its variable nature, many different distance calculations have been made for W\,Aql \citep{Danchi1994,Groenewegen1998,Tatebe2006,Ramstedt2009}. At a distance of 400 pc \citep{Whitelock2008}, this separation corresponds to 190 AU.

More recently, \cite{Mayer2013} used aperture photometry to classify the main sequence component of W\,Aql. They found a stable magnitude of $V\approx 14.8$, which  implies an absolute magnitude of $M_{\rm V}=7.1$ for an unreddened star at their adopted distance of $340$ pc, which would be consistent with a K4 dwarf. This is very different to \citeapos{Herbig1965} initial classification.

Previous studies have shown an extended dusty circumstellar envelope (CSE) around W\,Aql. In \citeapos{Ramstedt2011} observations of polarised visible light, the CSE extends more than 10\arcsec{} around the primary, which is well beyond the offset between the two binary components. This would certainly cause extinction depending on the alignment of the system. \citet{Tatebe2006} used the UC Berkeley Infrared Spatial Interferometer to resolve the distribution of emitting dust at 11.15 $\mic$ and found an asymmetry in the dust distribution around the primary on small scales (hundreds of mas), while \cite{Ramstedt2011} find asymmetry on large scales (10 arcseconds). \textit{Herschel}/PACS photometry presented in \cite{Mayer2013} shows a large ($\sim 100$ arcsecond) dust envelope on the same general shape as \cite{Ramstedt2011} found in the optical but with an bright patch of dust to the east of the star in both the 70 $\mic$ and 160 $\mic$ images.

Having a more precise classification of the main sequence companion allows us to better constrain several properties of the W\,Aql system. The mass of the main sequence component, which scales with spectral type, puts a lower limit on the initial mass of the AGB component; as the AGB component is more evolved, it must have had a larger initial mass. Further, the main sequence component imparts additional energy to the CSE of the AGB component, which could have an effect on the chemistry of the CSE, for example, through additional heating.

With conflicting results for the classification of the W Aql companion (F versus K spectral types), we investigate the binary system to find a more precise classification. We present new spectroscopic and photometric observations made during the minimum phase of the AGB component and new spectroscopic observations taken during the maximum phase of the AGB star for comparison. We present our observations in \sec{sec:obs}, perform the spectral analysis in \sec{sec:spec}, analyse the photometry in \sec{sec:phot}, and discuss the results in \sec{sec:dis}.

%__________________________________________________________________
\section{Observations and data reduction}\label{sec:obs}

%\subsection{Observations}

Observations were carried out using the Andalucia Faint Object Spectrograph and Camera (ALFOSC) on the Nordic Optical Telescope (NOT) on two nights in May 2012 and March 2013. The spectroscopy was done with a grism with 600 rules/mm, a dispersion of 1.5 \AA{}, and the wavelength range 3850--6850 \AA. The photometric observations were taken using UBV filters of the Bessel photometric system. The observations are summarised in Table \ref{tbl:obs}. The seeing on the night of our photometric observations ranged between 0\farcs5 -- 1\farcs1, so we were unable to resolve the two components of W\,Aql.

\begin{table}[tb]
\caption{Observations of W\,Aql.}             % title of Table
\label{tbl:obs}      % is used to refer this table in the text
\centering                          % used for centering table
\begin{tabular}{c c c c}        % centered columns (4 columns)
\hline               % inserts double horizontal lines
 Night of & Type & Configuration & Phase\\    % table heading 
\hline\hline                        % inserts single horizontal line
2012-05-11 & spectroscopy & 1\farcs0 slit, Grism \#7 & max\\
2013-03-29 & photometry & U\_Bes 362\_60 & min\\
2013-03-29 & photometry & B\_Bes 440\_100& min\\
2013-03-29 & photometry & V\_Bes 530\_80& min\\
2013-03-29 & spectroscopy & 1\farcs0 slit, Grism \#7& min\\
\hline                                   %inserts single line
\end{tabular}
\tablefoot{The numerical codes for the photometric filters give the central wavelengths followed by the full width at half maximum (FWHM) in nm. Max and min in the phase column refer to the periods of maximum or minimum brightness of W Aql.}
\end{table}

%\subsection{Data reduction}

Data reduction was carried out using IRAF (v 2.16). For the spectroscopy, we used a HeNe lamp for the wavelength calibration.

The standard reference star PG1525-071, and two other stars (labelled A and C) in the same field\footnote{http://www.not.iac.es/instruments/stancam/photstd/pg1525.html} (\cite{Landolt1992}) were used for our photometric calibration. To compensate for the different spectral response functions between the standard UBV system and the ALFOSC system, we applied the corrections as advised in the ALFOSC manual\footnote{http://www.not.iac.es/instruments/alfosc/zpmon/}. The integration times were 3$\times$200\,s, 20$\times$15\,s, and 10$\times1$\,s in U, B, and V, respectively. The standard stars were observed at air mass $\approx$1.24 and W\,Aql at air mass $\approx$1.66, and we used the standard extinction 0.46, 0.22, and 0.12 magnitudes/airmass in U, B, and V, respectively, to compensate for the differential atmospheric extinction. The U filter has small spectral leaks in the red spectral region beyond 7000\,\AA\footnote{see http://www.not.iac.es/instruments/filters/curves/png/7.png}, and for extremely red sources, like W\,Aql, there is a risk for contamination. However, a possible red ghost image would be displaced by 2$\farcs$5 at the zenith distance of the observations, $\approx$53$\degr$, and we can exclude a red ghost that is brighter than 0.5\,\% of the brightness in the U band.

%To better calibrate the B and V band magnitudes, reference stars in the W\,Aql field were identified for which previously observed magnitudes exist. We used the Naval Observatory Merged Astrometric Dataset (NOMAD) catalogue, which collates magnitude observations. The NOMAD identifiers for the field stars are given in Table \ref{refss} and the stars are labelled in Fig. \ref{fieldstars}.
%
%
%\begin{table}[tb]
%\caption{Reference stars in the W\,Aql field.}             % title of Table
%\label{refss}      % is used to refer this table in the text
%\centering                          % used for centering table
%\begin{tabular}{c c c c}        % centered columns (4 columns)
%\hline               % inserts double horizontal lines
%Our star ID & NOMAD ID & B Mag & V Mag \\    % table heading 
%\hline\hline                        % inserts single horizontal line
%A & 0829-0730521 & 14.470 & 13.890\\
%B & 0829-0730554 & 13.660 & 13.130\\
%C & 0829-0730577 & 14.110 & 13.900\\
%D & 0829-0730572 & 16.110 & 15.240\\
%E & 0829-0730639 & 13.000 & 12.430\\
%F & 0829-0730332 & 13.590 & 13.400\\
%\hline                                   %inserts single line
%\end{tabular}
%\end{table}
%
%   \begin{figure}
%   \centering
%   \includegraphics[width=8.8cm]{Bband-labelled.jpg}
%   \caption{W\,Aql and field Stars, B band.}
%              \label{fieldstars}%
%    \end{figure}

%____________________________________________
\section{Spectral analysis}\label{sec:spec}

When the AGB star is at maximum brightness, it completely dominates the spectrum at almost all wavelengths. The only aspect of the spectrum that might possibly show a contribution from the companion is a slight rise in the continuum in the blue from around 4200 \AA{} and bluewards.

When the AGB star is at minimum brightness, however, several features, which are clearly from the companion star, become visible in the blue end of the spectrum. The companion star dominates the spectrum up to around 5800--6000 \AA{}, redwards of which features that are characteristic of the AGB star again become apparent. Table \ref{features} summarises all the spectral features and the spectrum itself is shown in Fig. \ref{spec}.

\subsection{Features present only at minimum brightness}\label{minfeat}

Most of the features unique to the minimum brightness spectrum are found in the blue end, where the companion star dominates. The Balmer series is the most prominent feature with absorption lines of H$\beta$, H$\gamma$, and H$\delta$ present, which are absent in the maximum brightness spectrum. There are also strong Ca II K and H absorption features with the Ca II H line possibly blended with H$\varepsilon$. The strengths and shapes of these lines suggest that the companion star is an F star. The absence of H8 and H9 Balmer absorption lines bluewards of the Ca II pair suggests a later classification than F5 such as F8 { \citep{Gray2009}}.

Several less prominent metal lines are also present. The weak Ca I 4226 \AA{} line suggests an early to mid-F classification, as does the absence of the Fe I 4046 \AA{} line. The Mg I triplet around 5170 \AA{} is not stronger than the nearby H$\beta$ line, which suggests an early-F classification. The slightly more prominent Fe I 4383 \AA{} line suggests a mid-F classification. The weakness of these metal lines could be due to the star being relatively metal-poor. This would explain the apparent discrepancy between the metal lines and the Balmer series. The strength of the G band due to the CH molecule, which is comparable to H$\gamma$, also supports a later classification { \citep{Gray2009}}.

Overall, the spectral features suggest an F8 or F9 classification for the companion star and do not conclusively rule out G0.

\subsection{Features present only at maximum brightness}\label{maxfeat}

In stark contrast to the minimum brightness spectrum, the Balmer series lines H$\beta$ and H$\gamma$ are strongly in emission, indicating shocks in the AGB star. None of the absorption features discussed in \sec{minfeat} are present. 

There are strong ZrO bands throughout the spectrum but only a few generally weaker TiO bands, which suggests a relatively high C/O ratio for the S star ($\sim 0.98$, see Sec. \ref{agbprop}). There are also possible blendings of ZrO with YO and LaO (see Table \ref{features}).

\subsection{Features present during both phases}\label{bothfeat}

Mostly features at the red end of the spectrum are visible during both phases.

The sodium doublet (Na D) at 5892 and 5898 \AA{} is present for both phases but with a significant difference in shape. At minimum brightness, the absorption is narrower, and the two components of the doublet are clearly distinguished with a small peak between them. At maximum brightness, the Na D absorption is bordered by ZrO and YO and is much stronger with broader wings. Although the trough is slightly asymmetric, it is not possible to distinguish the two components. The wings of the Na D lines are apparent nearly up to the neighbouring YO band in the maximum brightness spectrum, while there is a clear gap between Na D and the same, albeit weaker, YO band in the minimum brightness spectrum. It is likely that the Na D feature comes from the AGB star in both cases, although it is possible there is some contribution from the F star at minimum brightness.

As for the above-discussed YO band at 5972 \AA, some redder ZrO bands in the 6300 to 6500 \AA{} range appear during both phases. The bands are much weaker in the minimum brightness spectrum, which is to be expected.

The final feature common to both phases is the H$\alpha$ emission line. As with the other Balmer emission lines discussed in \sec{maxfeat}, it most likely originates in shocks in the AGB star. In both cases, it is located on the side of a broad absorption trough, which is less deep at minimum brightness, making it difficult to compare with the other Balmer lines. \cite{Woodsworth1995} performed a detailed analysis of H$\alpha$ lines in S stars at different phases, but we do not have sufficient resolution in our spectrum to compare with his results.

\begin{table}[tb]
\caption{Spectral features at different phases.}             % title of Table
\label{features}      % is used to refer this table in the text
\centering                          % used for centering table
\begin{tabular}{l c c c}        % centered columns (4 columns)
\hline               % inserts double horizontal lines
Feature & $\lambda[\AA]$ & Maximum & Minimum \\    % table heading 
\hline\hline                        % inserts single horizontal line
H$\alpha$ 	& 6563				& emission & emission\\
H$\beta$		& 4861				& emission & absorption\\
H$\gamma$	& 4340				& emission & absorption\\
H$\delta$	& 4102				& ... & absorption\\
CaII H (+ H$\varepsilon$)	& 3970		& ... & absorption\\
CaII K		& 3934				& ... & absorption\\
Ca I 		& 4226				& ... & absorption\\
Fe I 		& 4383				& ... & absorption\\
MgI			& 5167, 5173, 5184	& ... & absorption\\
G band (CH)	& $\sim 4300$		& ... & present\\
Na D			& 5892, 5898			& absorption & absorption\\
%ZrO	bands	& many				& present & reduced\\
ZrO bands & 4620, 4641, 5304         & present & ...\\  
 	         & 5562, 5718, 5724	 	& present & ...\\
                  & 5849                                   & present & ... \\	         
        & 6136, 6344, 6474 & present & weaker\\ 
        & 6495	& present & weaker \\     
%TiO bands	& many				& weak & ... \\
YO bands		& 5972, 6132				& present & weak\\
\hline                                   %inserts single line
\end{tabular}
\tablefoot{References for wavelengths: \cite{Kramida2013} and \cite{Gray2009}.}
\end{table}

   \begin{figure*}
   \centering
   \includegraphics[width=8.8cm]{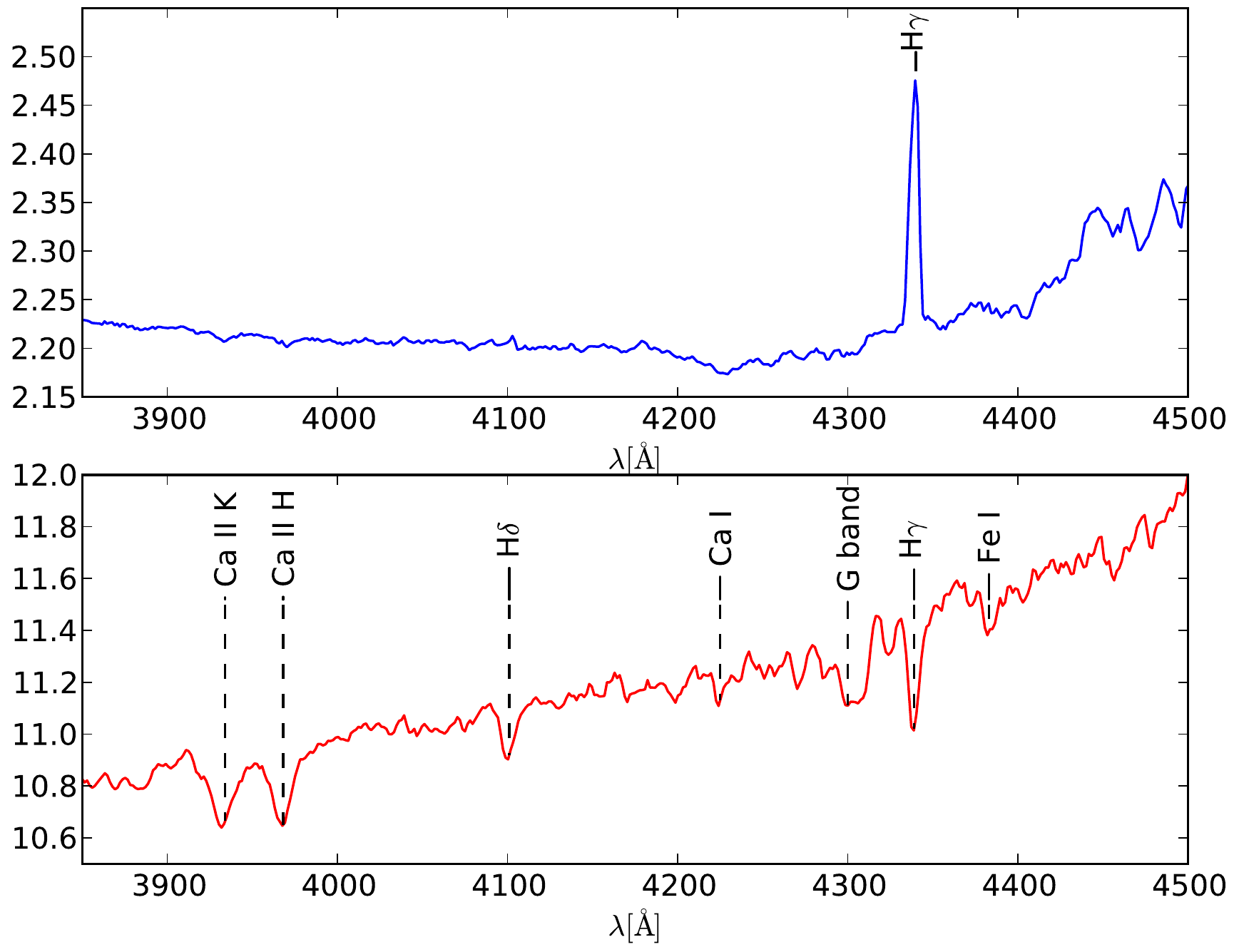}
   \includegraphics[width=8.8cm]{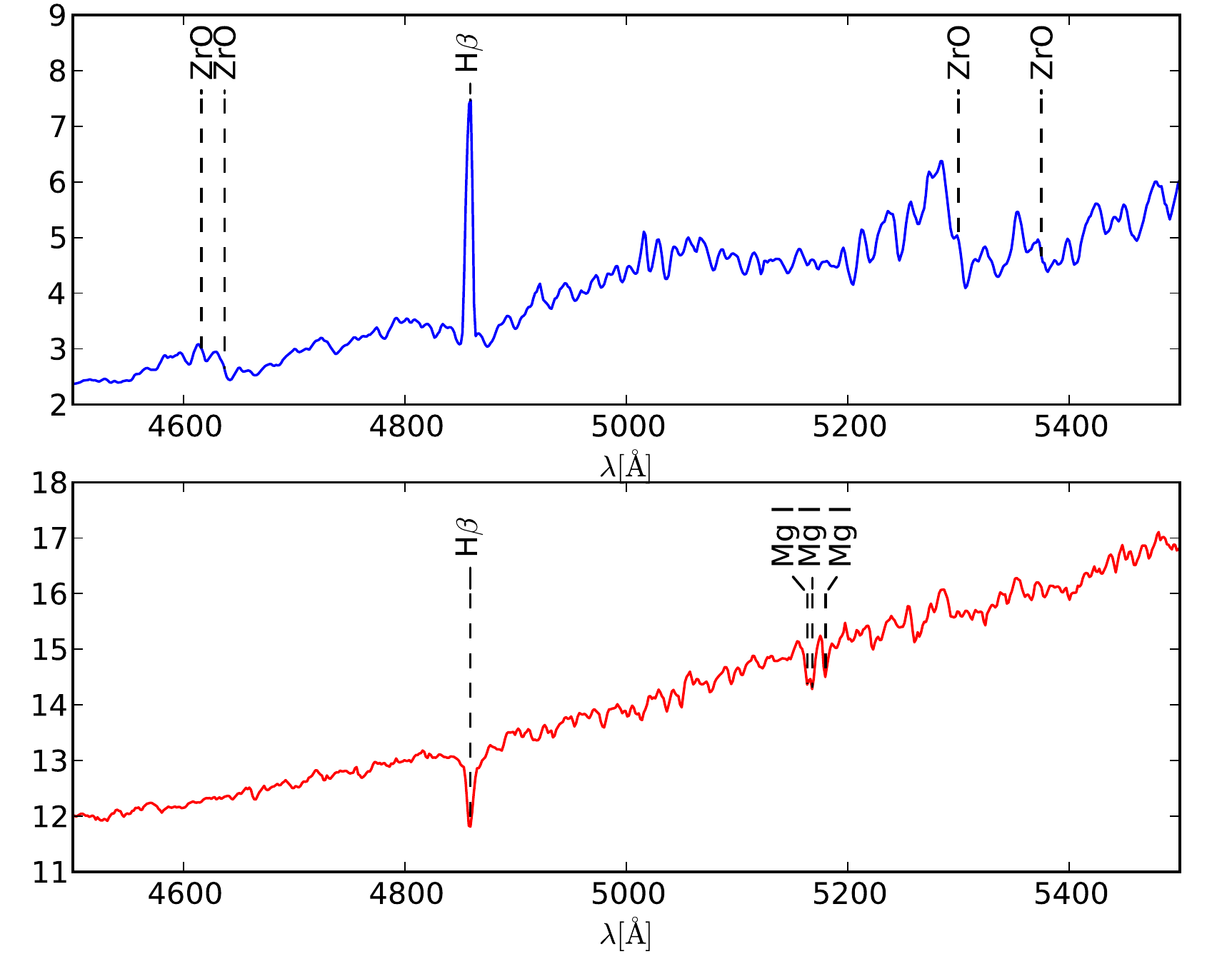}
   \includegraphics[width=8.8cm]{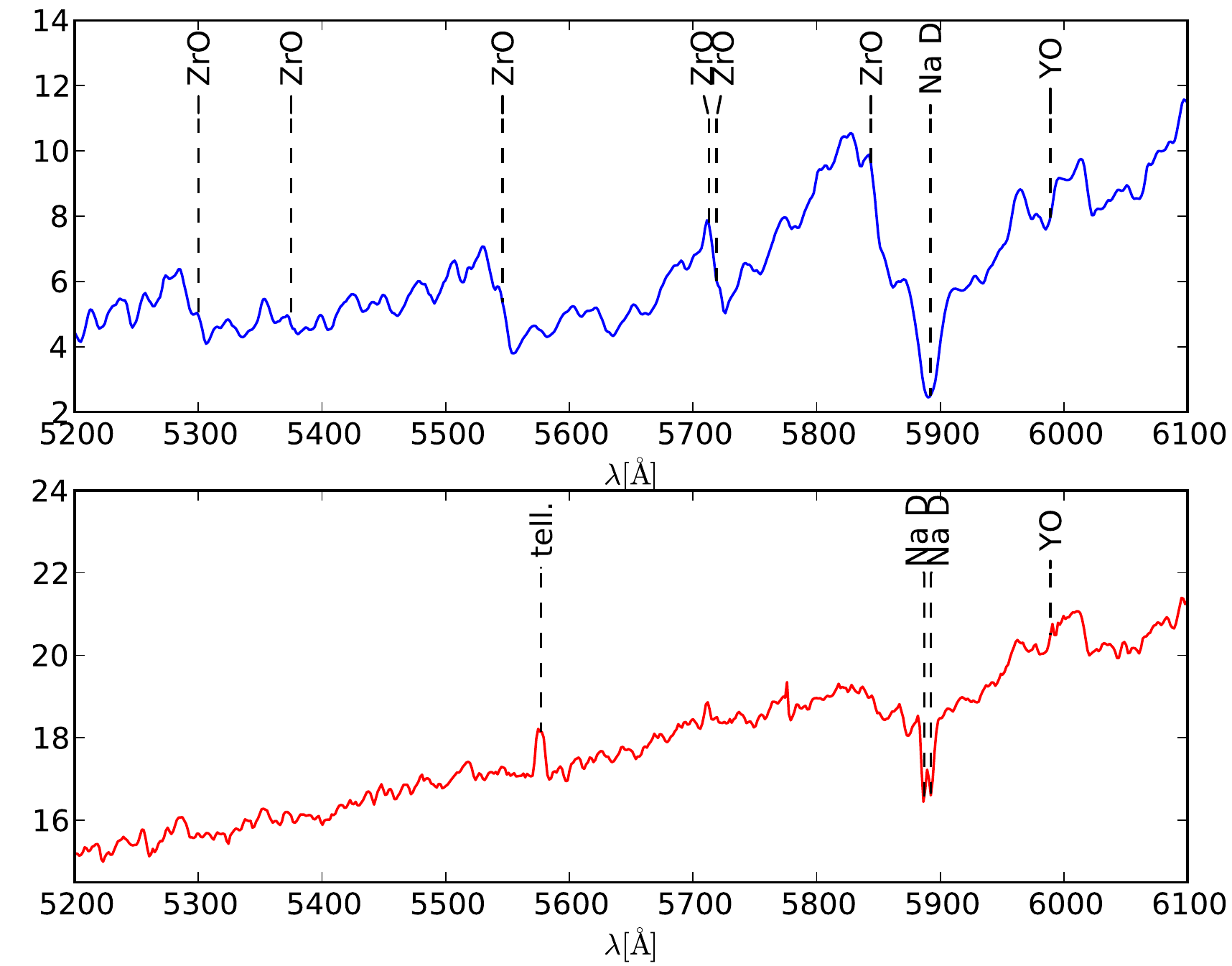}
   \includegraphics[width=8.8cm]{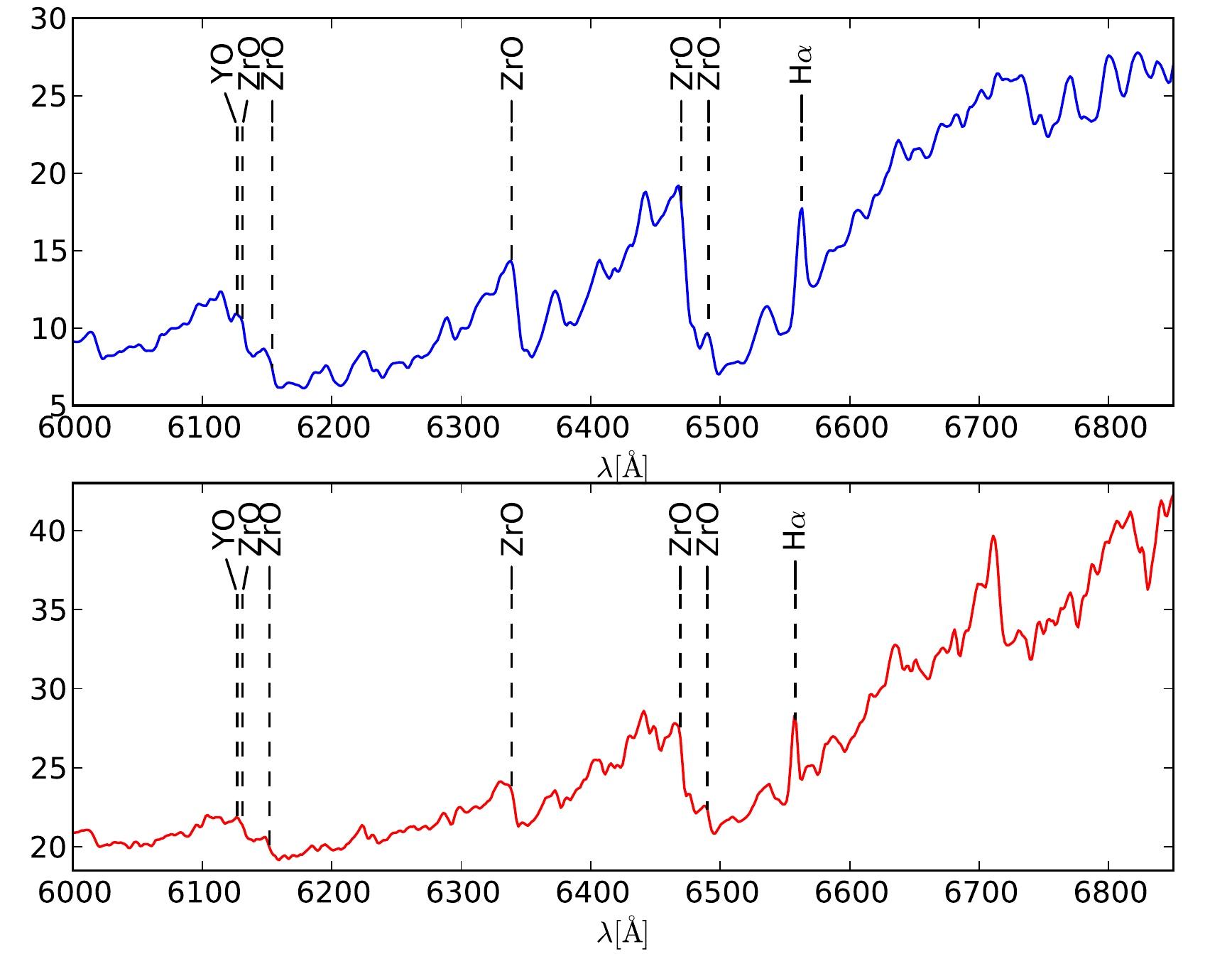}
   \caption{Observed spectra. The blue curve is the spectrum at maximum brightness and the red curve is the spectrum at minimum brightness. The vertical axes are in arbitrary flux units.}
              \label{spec}%
    \end{figure*}

\section{Photometric results}\label{sec:phot}

\begin{table}[tb]
\caption{Observed photometric results at phase minimum.}             % title of Table
\label{phot}      % is used to refer this table in the text
\centering                          % used for centering table
\begin{tabular}{c c c}        % centered columns (4 columns)
\hline               % inserts double horizontal lines
Band & Flux (mag) & Uncertainty\\    % table heading 
\hline\hline                        % inserts single horizontal line
U & 16.01 & 0.03\\
B & 15.26 & 0.01\\
V & 14.12 & 0.01\\
$U-B$ & \phantom{0}0.75 & 0.03\\
$B-V$ & \phantom{0}1.14 & 0.02\\
\hline
%$V_{\mathrm{corr}}$ & 14.17 & 0.03\\
%$(B-V)_{\mathrm{corr}}$ & \phantom{0}1.11 & 0.05\\
%M(U) & 8.02 & 0.09\phantom{0}\\
%M(B) & 7.27 & 0.02\phantom{0}\\
%M(V) & 6.09 & 0.015\\
%\hline%\hline
%Quantity & F8 value & G0 value\\
%\hline
%$A_U$ & 3.08 & 2.98\\
%$A_B$ & 2.75 & 2.29 \\
%$A_V$ & 2.09 & 1.69\\
%$E(U-B)$ & 0.73 & 0.69\\
%$E(B-V)$ & 0.66 & 0.60\\
%$\frac{E(U-B)}{E(B-V)}$ & 1.11 & 1.15\\
%$R_V$ & 3.17 & 2.82\\
%\hline                                   %inserts single line
\end{tabular}
%\tablefoot{The bands referenced here refer to the Bessel photometric system, as per our observations. The uncertainty cited is due to calibration; the stability of our measurements is good. The lower two values are corrected to be the contributions from the fainter star. See text.}
\end{table}

%\begin{table}[tb]
%\caption{Canonical photometric values for dwarf stars.}             % title of Table
%\label{canphot}      % is used to refer this table in the text
%\centering                          % used for centering table
%\begin{tabular}{l c c}        % centered columns (4 columns)
%\hline              % inserts double horizontal lines
%F8 & $M_V$& 4.0\\
%   & $(U-B)_0$ & 0.02\\
%   & $(B-V)_0$ & 0.52\\
%\hline
%F9 & $M_V$& 4.2\\
%\hline
%G0 & $M_V$& 4.4\\
%   & $(U - B)_0$ & 0.06\\
%   & $(B - V)_0$ & 0.58\\
%\hline                                   %inserts single line
%\end{tabular}
%\tablefoot{Quantities taken from \cite{Cox2000} and \cite{Gray2009}.}
%\end{table}

Our photometric results are summarised in Table \ref{phot}, which includes the $U-B$ and $B-V$ colours.
It should be noted that our photometric results are not exclusively of the fainter companion and include contamination by the AGB star, as the seeing was insufficient to resolve the two components of W\,Aql. 

Based on our observations, we can make some further deductions. If we assume that there is no extinction, which is an unlikely prospect given the dusty AGB envelope \citep[see][Fig. 2]{Danilovich2014}, we find that the absolute magnitude, assuming a distance of 400 pc, is $M_{V,\mathrm{obs}} = 6.1$ mag in the V band. This would suggest a K1 classification \citep{Gray2009}, similar to what \cite{Mayer2013} found. However, if we instead assume that our spectral classification is accurate, we can use canonical values \citep[taken from][]{Cox2000,Gray2009} to calculate the extinction. Considering $(B-V)_0 = 0.52$ for an F8V star, the excess $E(B-V)=0.62$ is consistent with an extinction of $A_V = 1.9$ mag, assuming standard interstellar extinction \citep[$R_V = 3.1$,][]{Draine2011}. If we calculate the extinction based on $M_{V\mathrm{,F8V}} = 4.0$ instead, we find $A_V = M_{V,\mathrm{obs}} - M_{V\mathrm{,F8V}} = 2.1$ mag. This could suggest a slight difference in the dust being produced by the AGB star as compared with dust in the interstellar medium (ISM).
{ Another source of uncertainty in the derivation of $A_V$ occurs because the $V$ and $B-V$ measurements are not exclusively of the F star but are likely contaminated by the AGB star.}

%______________________________________________
\section{Discussion}\label{sec:dis}

\subsection{Orientation of system}

The AGB component is surrounded by a dusty envelope and the envelope extent is significantly larger than the apparent separation between the AGB star and its companion \citep{Ramstedt2011,Mayer2013}. The significant extinction we calculate of $A_V = 2$ suggests that the companion star most likely sits within or behind the AGB star's CSE. Although some of the extinction we see may be interstellar, we believe it is unlikely to account for all the extinction, particularly as no interstellar contamination has been seen in W\,Aql's molecular emission lines \citep{Danilovich2014,Ramstedt2009,Schoier2013}.

\subsection{Properties of the fainter companion}

\citet{Mayer2013} used a luminosity-based classification for the companion. Using HST images, they calculated a K4V classification, which is significantly later than our classification of F8 or F9. We believe the difference arises in the lack of consideration of dust in \citeapos{Mayer2013} analysis.

Spectrally, the absence of the MgH feature at 4780 \AA{} and the MgH band at 5198 \AA{} in our spectrum (see Fig. \ref{spec}) counters the K4 classification, as does the presence of the Balmer absorption lines, the strength of the Ca II K and H absorption lines, and the strength of the G band, which should be much shallower by the mid-K.

Our result of an F8 to G0 star gives an effective temperature in the range of 6170 -- 5900 K \citep{Gray2009} and a stellar mass of 1.09 -- 1.04 $\msol$ \citep{Habets1981}.

%   \begin{figure}
%   \centering
%   \includegraphics[width=8.8cm]{AAVSO.pdf}
%   \caption{Long range light curve for W\,Aql from 1912 to 2012.}
%              \label{AAVSO}%
%    \end{figure}

\subsection{Properties of the AGB component}\label{agbprop}

The most prominent spectral features of the AGB component are the multiple ZrO bands, some of which are still visible in the red end of the spectrum even at the AGB star's minimum brightness. On the other hand, no clear TiO bands are seen in either minimum or maximum spectra. In general, TiO molecules are less resistant to temperature increases than ZrO and hence TiO bands are expected to be weak at maximum brightness and potentially more prominent at minimum brightness \citep{Richardson1933}. Their absence could be due to the companion dominating the spectrum, but this could only be true at the bluer end. The absence of TiO at the redder end of the spectrum, even at minimum brightness, indicates that the AGB component is more C-like than M-like in terms of the C/O abundance index. Following the criteria laid out in \cite{Keenan1980}, we find C/O $\sim 0.98$ with the classification of S6/6e (the first 6 being the temperature classification, taken from \citeapos{Keenan1980} observational results of W\,Aql and the e indicating the presence of emission lines). This is the highest C/O ratio before the SC classification takes over and is broadly consistent with molecular abundances determined in \citet{Danilovich2014}.

If we assume that the two components of W\,Aql formed contemporaneously, before it left the main sequence, the AGB star must have been more massive than the fainter companion. Additionally, the presence of s-process elements and more specifically Tc \citep{Little-Marenin1988} constrains the AGB star's mass to less than $\sim3$ $\msol$ \citep{Herwig2005}. This allows us to constrain the total system mass to 2.1 -- 4.1 $\msol$.% which is in line with \citeapos{Mayer2013} results using evolutionary tracks and a lower mass for the MS component.

\section{Conclusions}

The main conclusions of this paper are as follows:

   \begin{itemize}
      \item Our spectroscopic observations of W\,Aql suggest that the fainter component of the binary star is an F8 or F9 star due to the strength of the Ca II H and K lines, the Balmer absorption lines, and the Ca I, Fe I, and Mg I lines.
%      \item { We confirm that the companion star is physically associated to the AGB star by analysing scattered light around it.}
      \item The strength of the ZrO bands and lack of TiO bands in our spectral observations indicates that the AGB component of W\,Aql is an S6/6e star, which agrees with previous results.
      \item Our photometric results taken at minimum light are consistent with a star of luminosity class V at a distance of 400 pc and extinction $A_V = 2$  mag. \citet{Mayer2013} mistakenly attributed a spectral type of K4 because they neglected the large correction for extinction.
      %\item { W\,Aql exhibits large brightness variations at minimum light and we propose that this is caused by a patchy outflow. Dedicated photometric and spectroscopic observations at minimum brightness would allow characterisation of the optical properties of the dust as well as the structure of the outflow.}
      \item We are able to constrain the mass of the AGB star to \mbox{1.04 -- 3} $\msol$ based on the classification of the fainter star. This constrains the total system mass to 2.1 -- 4.1 $\msol$.
   \end{itemize}

\begin{acknowledgements}
      TD and KJ acknowledge funding from the SNSB.     
	  Based on observations made with the Nordic Optical Telescope, operated by the Nordic Optical Telescope Scientific Association at the Observatorio del Roque de los Muchachos, La Palma, Spain, of the Instituto de Astrofisica de Canarias.
	  We acknowledge with thanks the variable star observations from the AAVSO International Database contributed by observers worldwide and used in this research.
	  This research has made use of the International Variable Star Index (VSX) database, operated at AAVSO, Cambridge, Massachusetts, USA.
\end{acknowledgements}

%\bibliographystyle{../aa}
%\bibliography{../../Papers/AGBpapers.bib}

\end{document}